\documentclass[%
 reprint,
superscriptaddress,
nofootinbib,
 amsmath,amssymb,
 aps,
]{revtex4-2}
\usepackage{graphicx}% Include figure files
\usepackage{dcolumn}% Align table columns on decimal point
\usepackage{bm}% bold math
\usepackage[version=4]{mhchem} %chemical elements
\usepackage{subfig}
\usepackage{float}
\usepackage{caption}
\usepackage{multirow}
\usepackage[colorlinks,
            linkcolor=red,
            anchorcolor=blue,
            citecolor=blue
           ]{hyperref}% add hypertext capabilities
\graphicspath{ {figure/} }
\begin{document}

\title{Uncertainty in Hadronic Diffuse $\gamma$-Ray Emission \\ from the Temporal Stochasticity of Cosmic-Ray Sources}
%Slow-Diffusion Disk model}% Force line breaks with \\
%\thanks{A footnote to the article title}%

\author{Xing-Jian Lv}
\email{lvxj@ihep.ac.cn}
 \affiliation{%
 Key Laboratory of Particle Astrophysics, Institute of High Energy Physics, Chinese Academy of Sciences, Beijing 100049, China}
\affiliation{
 University of Chinese Academy of Sciences, Beijing 100049, China 
}%
 \author{Xiao-Jun Bi}
 \email{bixj@ihep.ac.cn}
\affiliation{%
 Key Laboratory of Particle Astrophysics, Institute of High Energy Physics, Chinese Academy of Sciences, Beijing 100049, China}
\affiliation{
 University of Chinese Academy of Sciences, Beijing 100049, China 
}%
\author{Kun Fang}
\email{fangkun@ihep.ac.cn}
\affiliation{%
 Key Laboratory of Particle Astrophysics, Institute of High Energy Physics, Chinese Academy of Sciences, Beijing 100049, China}
 \author{Han-Xiang Hu}
\email{hanxiang.hu@ihep.ac.cn}
 \affiliation{%
 Key Laboratory of Particle Astrophysics, Institute of High Energy Physics, Chinese Academy of Sciences, Beijing 100049, China}
\affiliation{
 Tibet University, Lhasa 850000, China 
}%
 \author{Peng-Fei Yin}
\email{yinpf@ihep.ac.cn}
\affiliation{%
 Key Laboratory of Particle Astrophysics, Institute of High Energy Physics, Chinese Academy of Sciences, Beijing 100049, China}
\author{Meng-Jie Zhao}
\email{zhaomj@impcas.ac.cn}
 \affiliation{%
 State Key Laboratory of Heavy Ion Science and Technology, Institute of Modern Physics, Chinese Academy of Sciences, Lanzhou 730000, China}

% \homepage{http://www.Second.institution.edu/~Charlie.Author}
%\affiliation{
% Third institution, the second for Charlie Author
%}%
%\author{Delta Author}
%\affiliation{%
% Authors' institution and/or address\\
% This line break forced with \textbackslash\textbackslash
%}%

%\collaboration{CLEO Collaboration}%\noaffiliation

\date{\today}% It is always \today, today,
             %  but any date may be explicitly specified

\begin{abstract}
Diffuse $\gamma$-ray emission is a key probe of cosmic rays (CRs) distribution within the Galaxy. However, the discrepancies between observations and theoretical model expectations highlight the need for refined uncertainty estimates. 
In the literature, spatial and temporal variability of lepton flux has been discussed as an uncertainty in diffuse $\gamma$-ray estimation.
In the present work, we demonstrate that 
variability in the high energy CR hadron flux is an important, yet previously underappreciated, source of uncertainty in diffuse $\gamma$-ray estimates.
To assess this effect, we perform fully three-dimensional, time-dependent GALPROP simulations of CR protons injected from discrete Galactic sources. Our results reveal that the uncertainty in the hadronic component of diffuse $\gamma$ rays is non-negligible and can be comparable to, or even exceed, current experimental uncertainties at very high energies. This finding challenges the conventional assumption that only leptonic fluctuations are relevant to diffuse $\gamma$-ray modeling.
\end{abstract}

%\keywords{Suggested keywords}%Use showkeys class option if keyword
                              %display desired
\maketitle

%\tableofcontents

\section{\label{sec:level1}INTRODUCTION}
Diffuse $\gamma$-ray emission is a key probe of Galactic cosmic rays (GCRs), originating from interactions of high-energy particles with the interstellar medium (ISM) and the interstellar radiation field (ISRF)~\cite{Tibaldo:2021viq}. Produced through both hadronic and leptonic processes, this emission encodes crucial information on the acceleration and propagation of GCRs across the Milky Way (MW)~\cite{BeckerTjus:2020xzg}. 
Recent advances in $\gamma$-ray astronomy, particularly at very high energies (VHE; $E_\gamma \gtrsim 100$~GeV), have significantly enhanced the precision and energy reach of diffuse $\gamma$-ray measurements~\cite{Fermi-LAT:2012edv, HESS:2014ree, ARGO-YBJ:2015cpa, LHAASO:2024lnz}. Despite these advances, growing tension between observational data and theoretical predictions---particularly in the VHE regime---highlights the need for more refined theoretical modeling of diffuse $\gamma$ rays, including more comprehensive understanding of theoretical uncertainties~\cite{Zhang:2023ajh}.

Among the various sources of theoretical uncertainty, increasing attention has been paid to the intrinsic stochasticity of cosmic ray (CR) sources, which introduces unavoidable variance in both the locally measured CR flux and the resulting diffuse $\gamma$-ray emission at Earth~\cite{Porter:2019wih, Marinos:2024rcg, Marinos:2023cgg}. This stochastic nature arises from the discrete, transient, and spatially inhomogeneous distribution of individual CR accelerators in the MW. Previous studies have primarily focused on the leptonic component of this effect, motivated by the fact that CR electrons and positrons suffer severe energy losses---via synchrotron radiation and inverse Compton scattering---on relatively short timescales, especially at very high energies~\cite{Moskalenko:1997gh}. This results in pronounced temporal and spatial fluctuations of the lepton flux in the MW, and consequently in its contribution to the diffuse $\gamma$-ray background. In contrast, hadronic CRs are considerably more stable, with energy loss timescales that greatly exceed their confinement times~\cite{Strong:1998pw}. Furthermore, because diffuse $\gamma$-ray signals are measured through line-of-sight integration over the Galactic volume, it has long been assumed that fluctuations in the hadronic contribution are effectively averaged out and therefore negligible compared to current experimental uncertainties~\cite{Porter:2019wih, Marinos:2024rcg}.

However, previous studies have largely overlooked another important source of uncertainty arising from the intrinsic stochasticity of CR sources: the normalization of the CR flux. This normalization is  inferred from measurements taken at a single point in space and time—namely, at the Solar System today—but is highly sensitive to the recent local source history over the past few million years. If one or more nearby CR sources have been active during this period, the locally measured CR flux, and thus the inferred normalization, will exceed the Galactic average; conversely, the absence of recent nearby sources yields a lower value. Crucially, this type of uncertainty is not mitigated by line-of-sight integration in the diffuse $\gamma$-ray signal, as it directly propagates from the CR proton normalization to the hadronic $\gamma$-ray component. In this work, we demonstrate that this contribution to the theoretical uncertainty is not only non-negligible but can be comparable to, or even exceed, current experimental uncertainties, particularly at very high energies.

This work is organized as follows. In Section~\ref{sec methology}, we describe 
the specific details of time-dependent modeling of CR propagation in the MW. Our results are presented in section~\ref{sec results}. In section~\ref{sec:conclusion}, we give a summary of our findings

\section{Modeling Setup}\label{sec methology}
The transport of GCRs is primarily modeled as a diffusion process, supplemented by additional physical processes such as reacceleration, energy losses, and fragmentation, and is described by~\cite{Strong:2007nh}:
\begin{equation}
\begin{aligned}
\frac{\partial \psi}{\partial t}= & Q(\mathbf{x}, p)+\nabla \cdot\left(D_{x x} \nabla \psi-\mathbf{V}_c \psi\right)+\frac{\partial}{\partial p}[p^2 D_{p p} \frac{\partial}{\partial p}(\frac{\psi}{p^2})] \\
& -\frac{\partial}{\partial p}[\dot{p} \psi-\frac{p}{3}(\nabla \cdot \mathbf{V}_c) \psi]-\frac{\psi}{\tau_f}-\frac{\psi}{\tau_r}\;, \label{eqn: transport}
\end{aligned}
\end{equation}
where $\psi(\mathbf{x}, p, t)$ is the CR density per unit momentum, and $Q(\mathbf{x}, p)$ denotes the CR source term. The coefficients $D_{xx}$ and $D_{pp}$ represent spatial and momentum diffusion, respectively, with $D_{xx}$ typically modeled as a rigidity-dependent power law. The momentum diffusion coefficient $D_{pp}$ is related to $D_{xx}$ via the relation $D_{pp} D_{xx} = 4 p^2 v_A^2/3 \delta (4 - \delta^2)(4 -\delta)$~\cite{seoStochasticReaccelerationCosmic1994,  berezinskiiAstrophysicsCosmicRays1990a}, where $v_A$ is the Alfv\'en speed and $\delta$ is the spectral index characterizing the rigidity dependence of $D_{xx}$. The convection velocity is denoted by $\mathbf{V}_c$, and $\dot{p} \equiv dp/dt$ is the momentum loss rate. The terms $\tau_f$ and $\tau_r$ correspond to the characteristic timescales for fragmentation and radioactive decay. 

We solve Eqn.~\ref{eqn: transport} with the numerical package GALPROP\footnote{\url{https://galprop.stanford.edu/}}~\cite{Strong:1998pw,Moskalenko:1997gh}, a widely used framework for CR propagation. In this work, we employ the latest public release (version~57), which features improved support for three-dimensional modeling and time-dependent solutions compared with earlier versions~\cite{Porter:2021tlr}.
For the GALPROP setup, we adopt the SA100 model distributed with version~57 (available in the \textit{example} directory). A full description can be found in Ref.~\cite{Porter:2021tlr}, which we briefly summarize below.

ISM distributions with spiral-arm structures consistent with observations are adopted.
The ISRF follows the \textit{R12} model~\cite{Robitaille:2012kg}, while the Galactic magnetic field is described by the \textit{Pshirkov\_BSS} model~\cite{Pshirkov:2011um}. The three-dimensional gas distribution is taken from  Ref.~\cite{Johannesson:2018bit}. CR propagation parameters are adopted from Ref.~\cite{Porter:2021tlr}, where they were calibrated against AMS–02, ACE/CRIS, and Voyager~1 data compiled in Ref.~\cite{Johannesson:2019jlk}. The force-field approximation~\cite{Gleeson:1967juf, Gleeson:1968zza} is used for Solar modulation, though its effect is negligible here since the analysis focuses on CR energies above tens of GeV~\cite{Potgieter:2013pdj}.  
The \textit{AAfrag} package\footnote{\url{https://sourceforge.net/projects/aafrag/}}~\cite{Kachelriess:2019ifk, Kachelriess:2022khq} is employed to compute $\gamma$-ray production from \textit{pp} collisions. Pair absorption on ISRF~\cite{Robitaille:2012kg} is taken into account, and the resulting skymaps are generated with seventh-order HEALPix~\cite{Gorski:2004by}.

The source term $Q(\mathbf{x}, p)$ is factorized into a spatially dependent part and an energy-dependent part. For the spatial distribution, we adopt the SA100 model~\cite{Porter:2017vaa}, which assumes a pure spiral-arm configuration, and randomly generate individual CR sources accordingly in the MW. For the energy-dependent part, the injection spectrum is parameterized as a broken power law in rigidity, and all sources share the same spectrum. To reduce computational cost, only protons are injected in the simulations, as they dominate the hadronic contribution to the diffuse $\gamma$-ray emission~\cite{Porter:2019wih} and the uncertainties in the emission from heavier nuclei are expected to be comparable. This choice reflects the focus of the present work, which is to estimate the uncertainty in diffuse $\gamma$-ray emission arising from temporal variations in CR densities, rather than to provide a full prediction of the diffuse $\gamma$-ray spectrum.

\begin{figure*}[htbp]
    \begin{center}
        \subfloat{\includegraphics[width=0.49\textwidth]{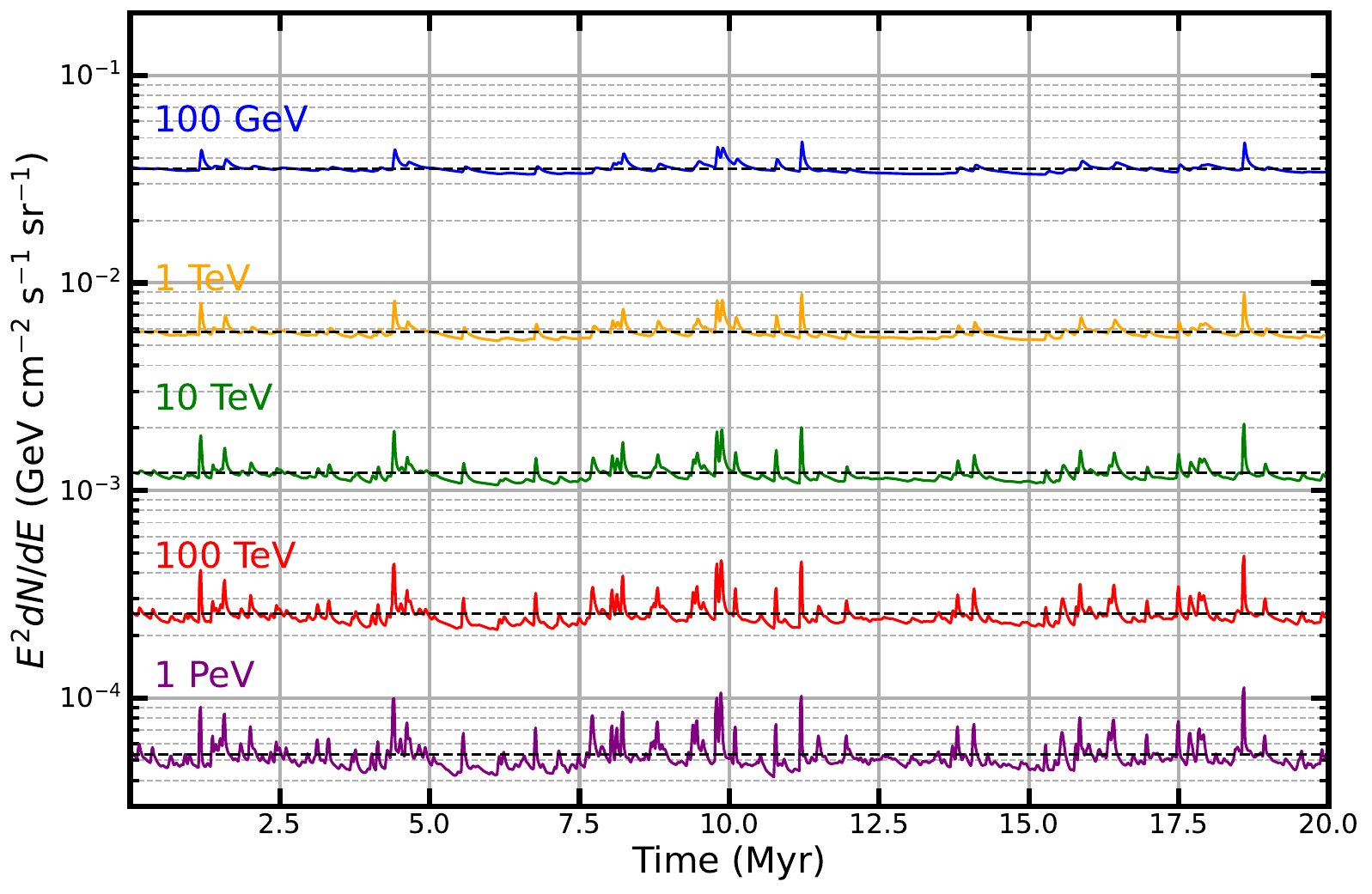}} \hskip 0.01\textwidth
        \subfloat{\includegraphics[width=0.47\textwidth]{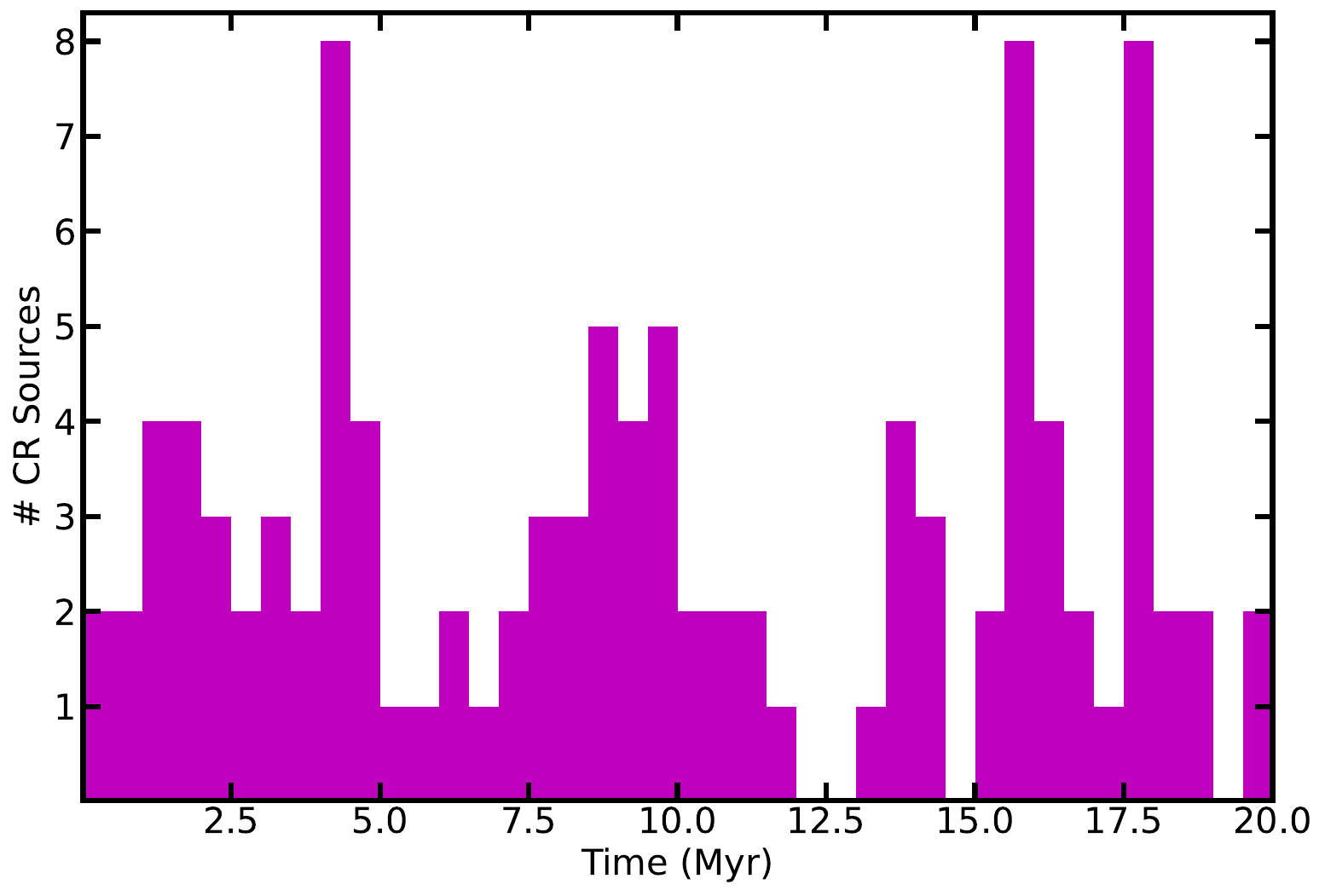}}
    \end{center}
    \captionsetup{justification=raggedright}
    \caption{Left: Temporal evolution of the CR proton flux at the solar location for selected energies (100~GeV, 1~TeV, 10~TeV, 100~TeV, and 1~PeV). Dashed black lines indicate the corresponding steady-state fluxes at each energy.  
    Right: Number of active CR sources within 1~kpc of Earth over the same time interval.}\label{fig: time flux}
\end{figure*}
\begin{center}
    \textit{Time Dependent Calculations}
\end{center}
For the time-dependent calculation, we adopt the two-step procedure described in Refs.~\cite{Porter:2019wih, Marinos:2024rcg, Marinos:2023cgg}: an initial phase during which the CR density approaches equilibrium, followed by a final propagation period lasting several Myr. The only deviation from this procedure is that the initial phase does not begin with a zero CR density throughout the MW, but instead is initialized using CR densities obtained from a steady-state calculation\footnote{This is the default solution method of GALPROP, which starts 
with a large initial time step ($\textit{start\_timestep} = 10^9$~yr) corresponding
to the longest timescales of the system. The timestep is then reduced iteratively
by a constant factor ($\textit{timestep\_factor} = 0.5$) until the final timestep ($\textit{end\_timestep} = 10$~yr)
is reached, which should be as short as the shortest timescales of the system.} with identical configuration settings. This approach reduces the time required to reach equilibrium from $\sim$100\,Myr to a few~kyr, thereby improving computational efficiency. Specifically, our simulation uses a timestep of $\Delta t = 500$~yr, with the initial phase lasting 0.5~Myr and the final phase 20~Myr, and outputs generated every 5~kyr for CR proton densities and $\pi^0$-decay $\gamma$ rays. All results presented in Sec.~\ref{sec results} are based on the final phase. 

For the time-dependent solution, the source creation rate $\mathcal{R}$ and source lifetime $\tau$ play a key role in setting the energy-dependent amplitude of temporal variations, yet both remain poorly constrained.  

The acceleration and the subsequent injection of relativistic particles are not yet fully understood. Nevertheless, most studies indicate that the duration of efficient particle acceleration in supernova remnants (SNRs) is energy-dependent: at the highest energies (\(\gtrsim\)PeV), acceleration persists for only \(\lesssim 10^2\)~yr in rare favorable environments~\cite{Bell:2013kq,Gaggero:2017abc,Cristofari:2021jkl}; for multi-TeV particles it typically lasts \(\sim 10^2\)–\(10^3\)~yr before the maximum energy drops below the multi-TeV domain as the shock decelerates~\cite{Telezhinsky:2011aa,Schure:2013kya}; and for sub-TeV particles it can continue up to \(\sim 10^4\)~yr, albeit with decreasing efficiency and luminosity~\cite{Dermer:2012vk,Brose:2019spv}.  

Constraints on $\mathcal{R}$ are likewise uncertain.
Estimates based on the decay of \({}^{26}\)Al suggest one SN every 35–125~yr~\cite{2006Natur.439...45D}, while OB-star counts imply one every 200–250~yr~\cite{10.1093/mnras/staf083}. Even lower rates, about one SN every 500~yr, have been inferred from the cosmic-ray electron spectrum~\cite{Mertsch:2018bqd}.

In this work, we adopt a representative combination of $\mathcal{R} = 1/500$~yr$^{-1}$ and $\tau = 10^4$~yr. Since the amplitude of fluctuations primarily depends on the average number of active CR sources in the MW, $n = \mathcal{R}\tau$, smaller $n$ leads to larger fluctuations~\cite{Marinos:2024rcg}. The adopted combination therefore represents a conservative choice, particularly at the highest energies where CR acceleration lasts the shortest time.

\section{RESULTS \& DISCUSSION}\label{sec results}
\subsection{COSMIC-RAY VARIABILITY}
\begin{figure}[htbp]
\includegraphics[width=0.5\textwidth]{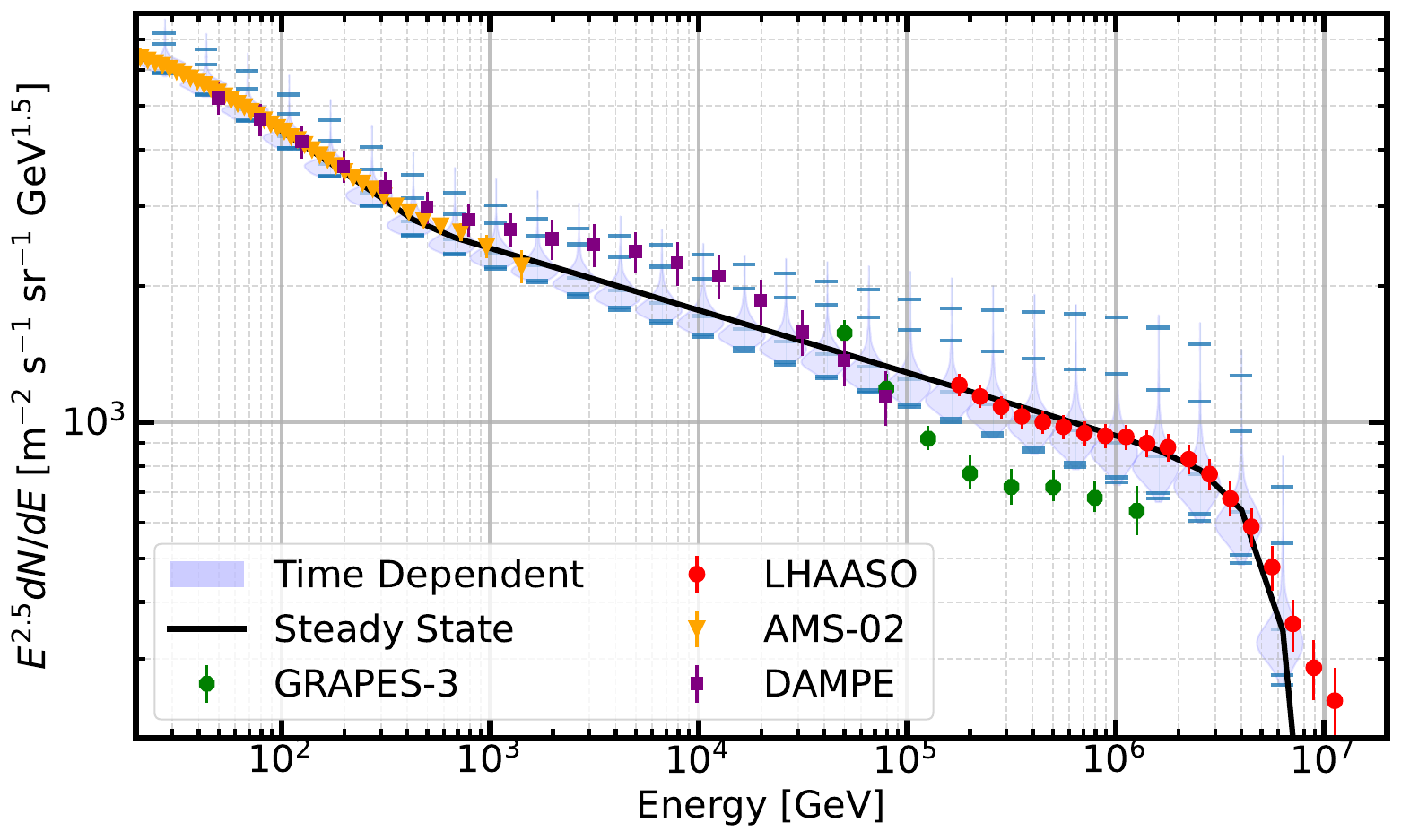}\\
\captionsetup{justification=raggedright}
\caption{Steady-state CR proton spectrum (solid black line) compared with the distribution from the time-dependent solution, shown as violin plots (light blue) for selected energy bins. Horizontal bars within each violin denote the mean (central line) as well as the 95\% and 99\% quantiles (inner and outer bars, respectively). Experimental data are included for reference: AMS–02~\cite{AMS:2021nhj}, LHAASO~\cite{LHAASO:2025byy}, DAMPE~\cite{DAMPE:2019gys}, and GRAPES–3~\cite{GRAPES-3:2024mhy}. \label{fig: mid}}
\end{figure}

Figure~\ref{fig: time flux} compares the time-dependent proton flux with the steady-state solution. 
The left panel presents the temporal evolution of the CR proton flux at selected energies (100~GeV, 1~TeV, 10~TeV, 100~TeV, and 1~PeV) at the solar location, along with the corresponding steady-state values (dashed black lines). 
The right panel shows the number of active CR sources within 1~kpc of the solar location as a function of time over the same interval.
Figure~\ref{fig: mid} compares the steady-state CR proton spectrum (solid black line) with the distribution obtained from the time-dependent solution, represented as violin plots (light blue) for selected energy bins. For reference, measurements from AMS–02~\cite{AMS:2021nhj}, DAMPE~\cite{DAMPE:2019gys}, GRAPES~3~\cite{GRAPES-3:2024mhy}, and LHAASO~\cite{LHAASO:2025byy} are included. 
Within each violin, the horizontal bars indicate the mean (central line) and the 95\% and 99\% quantiles (inner and outer bars, respectively).

The steady-state solution represents a density averaged over time, whereas the time-dependent solution exhibits fluctuations driven by the number of active CR sources within a few Myr of the solar system’s vicinity. The amplitude of these fluctuations increases with energy. Since the diffusive timescale scales as $t_d \propto 1/D(E)$, higher-energy particles escape more rapidly from the MW and are therefore more sensitive to the number and proximity of recent sources, making their flux dominated by a few nearby, young sources and consequently more variable.

As shown in the left panel of Fig.~\ref{fig: time flux}, large fluctuations are predominantly upward. An upward fluctuation can arise from the presence of just a single source located near the solar system, whereas a pronounced downward fluctuation would require the absence of nearby sources over an extended period, an event with very low probability. As shown in the inset, fluctuations in the CR density time series closely track the number of sources within 1~kpc of the solar system, with peaks in the CR density generally corresponding to higher local source counts. A similar behavior is evident in the violin plots in Fig.~\ref{fig: mid}, where the distributions deviate strongly from Gaussian, exhibiting long tails toward higher fluxes. These results are in perfect agreement with Ref.~\cite{Porter:2019wih}, although the fluctuations reported there are smaller due to the adoption of a longer source lifetime and a higher source rate.

\begin{figure}[htbp]
\includegraphics[width=0.5\textwidth]{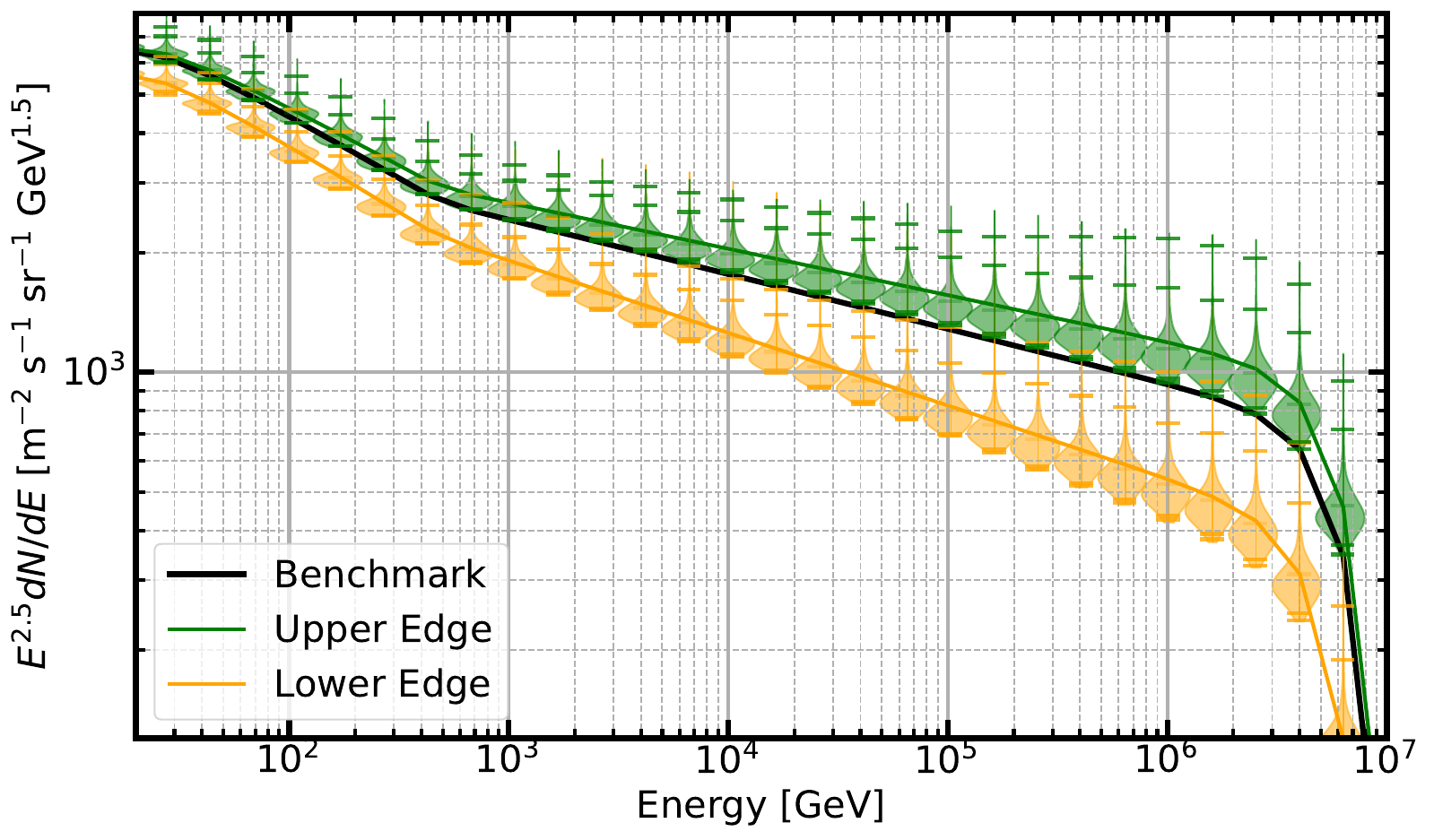}\\
\captionsetup{justification=raggedright}
\caption{The impact of local source history on the inferred CR proton flux. The solid black line shows the benchmark flux in Fig.~\ref{fig: mid}. The green and orange violin plots represent time-dependent solutions constrained to lie within the 99\% quantile of the distribution, corresponding to scenarios where the local CR flux is enhanced (Lower Edge) or suppressed (Upper Edge) relative to the Galactic average. \label{fig: lim}}
\end{figure}

Time-dependent fluctuations driven by the number of active CR sources within a few Myr of the solar system’s vicinity imply that the locally measured CR density may not reflect the Galactic average. If one or more sources are located nearby, the inferred CR density will be higher than the MW average; conversely, if no nearby sources have been active in the past few Myr, the inferred CR density will be lower. The extreme limits of these scenarios are illustrated in Fig.~\ref{fig: lim}, where the orange lines and violin plots correspond to the former case and the green to the latter. We adopt the steady-state CR flux in Fig.~\ref{fig: mid} as the benchmark CR flux at Earth, and in each case this benchmark (solid black line) is required to lie within the 99\% quantile of the time-dependent distribution. This choice reflects the non-Gaussian nature of the distribution, which exhibits long tails. The commonly used 95\% quantile may underestimate the range of realistic fluctuations, particularly at high energies where the variance is dominated by rare, nearby sources.

\subsection{GAMMA-RAY VARIABILITY}
\begin{figure}[htbp]
\includegraphics[width=0.5\textwidth]{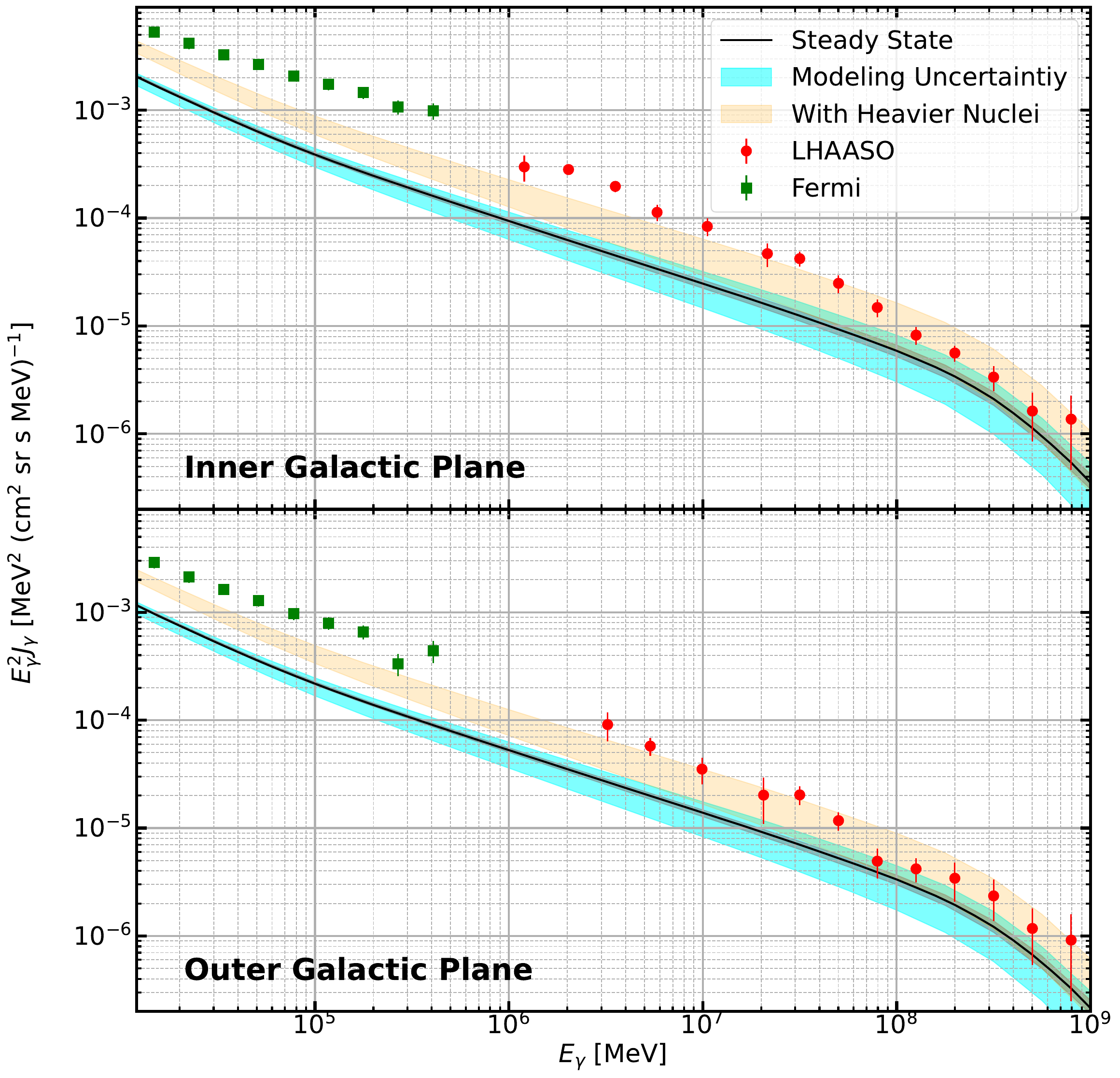}\\
\captionsetup{justification=raggedright}
\caption{Predicted $\gamma$-ray spectral uncertainty at Earth from the inner (top) and outer (bottom) Galactic plane, shown alongside Fermi-LAT (green) and LHAASO (red) measurements. All data and model predictions are processed using the same sky mask as in Ref.~\cite{LHAASO:2024lnz}. The solid black curves indicate the steady-state proton-induced emission, while the cyan bands represent theoretical uncertainties arising from source variability. The orange bands show an approximate enhancement by a factor of 2.0~\cite{Mori:2009te}, illustrating the potential contribution from heavier nuclei. The narrow gray bands indicate the much smaller theoretical uncertainties associated with variability in the number of sources along the line of sight through the MW.
\label{fig: gamma}}
\end{figure}

To quantify the theoretical uncertainty in the $\gamma$-ray spectrum at Earth induced by source variability, we evaluate three representative cases presented in Fig.~\ref{fig: lim}, with the corresponding results shown in Fig.~\ref{fig: gamma}. The solid black lines indicate the steady-state predictions based on the benchmark CR flux, while the cyan bands represent the associated theoretical uncertainty. The upper and lower panels correspond to emissions from the inner ($15^\circ < l < 125^\circ, \; |b| < 5^\circ$) and outer ($125^\circ < l < 235^\circ, \; |b| < 5^\circ$) Galactic plane, respectively.
To ensure direct comparability with the LHAASO observations~\cite{LHAASO:2024lnz}, we adopt the same sky mask as used in their analysis and apply it consistently to our model predictions. The Fermi-LAT data~\cite{Zhang:2023ajh} are likewise reprocessed with the same mask for consistency. Both LHAASO and Fermi-LAT measurements are overlaid in Fig.~\ref{fig: gamma} to highlight the relative scale of experimental uncertainties. Notably, the theoretical uncertainties in most energy bins are comparable to or exceed the corresponding observational error bars.
In particular, the theoretical uncertainty gradually increases from $\sim$30\% at 100~GeV to $\sim$100\% at 1~PeV in both the inner and outer Galactic plane.

It is important to emphasize that these observational data are not employed to calibrate or constrain the model. Our calculation includes only the hadronic contribution from protons, neglecting heavier nuclei and leptonic channels. For illustrative purposes, we scale the results by a factor of 2.0, following Ref.~\cite{Mori:2009te}, to approximate the potential contribution from heavier nuclei. This scaling is intended solely for illustration and is not rigorous, since the enhancement factor depends on the energy-dependent composition of both the ISM and CRs~\cite{Peron:2021uft}, and therefore carries its own uncertainties.

The uncertainty in the theoretical prediction arises from two sources. The first is the variability in the number of sources along the line of sight during the integration through the MW, represented by the extremely narrow gray bands in Fig.~\ref{fig: gamma}. This component is small, gradually increasing from $\lesssim 10\%$ at 100~GeV to $\sim 30\%$ at 1~PeV in both the inner and outer Galactic plane, since the total number of sources in the Galaxy is relatively stable, consistent with the findings of Ref.~\cite{Porter:2019wih}. For this reason, previous studies often neglected the uncertainties arising from temporal variability in the hadronic contribution to diffuse $\gamma$ rays. 

The second uncertainty comes from the overall normalization of the CR proton flux, which is actually the dominant uncertainty.
%contribution instead originates from the overall normalization of the CR proton flux. 
The normalization is determined from the flux measured at a single location and time---the Solar System today---and is highly sensitive to the local source history over the past few Myr. If one or more nearby sources have been active during this period, the inferred normalization will be higher than the Galactic average; conversely, an absence of such sources would yield a lower value. These scenarios, illustrated in Fig.~\ref{fig: lim}, directly map onto the range of $\gamma$-ray intensities shown by the cyan bands in Fig.~\ref{fig: gamma}.

In this work, the theoretical uncertainty shown in Fig.~\ref{fig: gamma} is calculated assuming a source creation rate of one every 500~yr and an active time of $10^4$~yr per source. A lower creation rate or shorter active time would increase the uncertainty, while higher values would reduce it. Our calculation further assumes that a single source population contributes to the Galactic CR flux. However, recent studies suggest that multiple classes of CR sources may exist, each dominating different energy ranges~\cite{Hoerandel:2002yg, Gaisser:2013bla, Lv:2024wrs}. Sources responsible for the highest energies tend to occur at lower rates and have shorter lifetimes~\cite{Cristofari:2021jkl}, which would enhance the uncertainty with increasing $\gamma$-ray energy and could give rise to the so-called “local knee” as proposed in Refs.~\cite{Prevotat:2024ynu, Prevotat:2025ktr}. 

In addition, recent advances in GCR propagation theory suggest that the diffusion coefficient may not be uniform throughout the halo; instead, CRs may diffuse more slowly near the Galactic disk~\cite{Tomassetti:2012ga, Guo:2018wyf, Zhao:2021yzf}. Such slow diffusion would increase the sensitivity of the local CR density to nearby source history~\cite{Yao:2023oay}, thereby enlarging the theoretical uncertainty in the predicted $\gamma$-ray flux. Finally, the leptonic contribution carries even larger uncertainties, as the normalization of the CR electron flux is far more sensitive than that of hadrons to the recent local source history, owing to its short energy-loss timescales.

\section{SUMMARY\label{sec:conclusion}}
We performed fully three-dimensional GALPROP simulations of time-dependent CR proton injection and propagation from discrete Galactic sources, and investigated their impact on the local CR density and diffuse $\gamma$-ray emission. In contrast to previous studies~\cite{Porter:2019wih}, which considered only the line-of-sight variability in the number of sources contributing to the $\gamma$-ray flux, our work also accounts for the uncertainty in the overall normalization of the CR flux. This normalization, inferred from measurements at a single point in space and time—the Solar System today—is highly sensitive to the stochastic nature of the local source history over the past few Myr. By including this additional source of uncertainty, we show that the resulting uncertainty in the hadronic contribution to the diffuse $\gamma$-ray flux is non-negligible and can be comparable to, or even exceed, current experimental uncertainties. This challenges the conventional assumption that only the variability in the leptonic contribution needs to be considered when modeling the diffuse $\gamma$-ray emission~\cite{Marinos:2024rcg}.

These findings highlight the importance of accounting for source-induced variability and local source history when modeling Galactic CRs and diffuse $\gamma$-ray emission. 
Future work should aim to refine these uncertainties by incorporating more realistic source populations with energy-dependent lifetimes and occurrence rates, and by investigating the impact of spatially varying diffusion in CR propagation models. 

\acknowledgments
This work is supported by the 
National Key R\&D program of China under the grant 2024YFA1611402, the National Natural Science Foundation of China under Grants No. 12175248, No. 12105292, and No. 12393853. 

\bibliography{apssamp}
%\appendix

% The \nocite command causes all entries in a bibliography to be printed out
% whether or not they are actually referenced in the text. This is appropriate
% for the sample file to show the different styles of references, but authors
% most likely will not want to use it.
%\nocite{*}
%
\end{document}